\documentclass[fp,twocolumn]{jpsj3}
\usepackage{txfonts}
\usepackage{bm}

\title{Impact of Softness of Particles on Rheology of Dilute Granular Gases}

\author{Haruto Ishikawa$^1$ and Satoshi Takada$^{1,2}$\thanks{Corresponding author. E-mail: takada@go.tuat.ac.jp}}
\inst{$^1$Department of Mechanical Systems Engineering, 
Tokyo University of Agriculture and Technology,\\
Koganei, Tokyo 184--8588 Japan \\
$^2$Institute of Engineering, 
Tokyo University of Agriculture and Technology, 
Koganei, Tokyo 184--8588 Japan}

\abst{
We numerically and theoretically investigate how the softness of particles affects the rheology of sheared dilute granular gases.
We find that the kinetic theory predicts the deviation of the flow curve from the Bagnold scaling, and it works well below a certain shear rate when we compare with the simulation results.
It is also found that there is no theoretical solution above this shear rate, which is because the energy loss due to inelastic collisions cannot be balanced with the energy injection by the shear.
}

\begin{document}
\maketitle

\section{Introduction}

The kinetic theory of hard-core granular gases is known to be a powerful tool to reproduce the rheology of dilute or moderately dense granular flows \cite{Brey98, Garzo99, Brilliantov, Garzo_book, Santos04, Montanero05, Mitarai07}.
For sheared systems, for example, the rheology is described by the Bagnold scaling, where the viscosity is proportional to the shear rate.
These results are also verified in terms of the direct simulation Monte Carlo \cite{Santos04, Montanero05} and event-driven molecular dynamics simulations \cite{Mitarai07}. 

Once the softness of particles is considered, on the other hand, another time scale due to this softness appears in the system.
The kinetic theory of soft-core gases has been constructed in many systems such as that having the inverse power law potential or Lennard--Jones potential \cite{Hirschfelder, Kihara43, Hollera51, Takada16, Takada18, Sanchez-Tena19}.
Recently, Sugimoto and Takada \cite{Sugimoto20} extended the kinetic theory to soft-core gases having the harmonic potential, where the collision angle is explicitly written in terms of the elliptic integrals.

In this paper, we numerically and theoretically investigate how the softness of particles affects the rheology of the system.
Numerically, we first attempt to model this system as the combination of the repulsive and the dissipative forces.
Then, we perform the molecular dynamics simulations under a shear.
Theoretically, we aim to extend the kinetic theory of frictionless soft-core gases to inelastic systems.
Here, the energy dissipation in the simulation is mapped onto the inelasticity in the kinetic theory.
We derive the temperature dependencies of the quantities from the Boltzmann equation.
Finally, we compare both results and check the applicability of our theoretical treatment.

The organization of this paper is as follows:
In the next section, we briefly explain the model and the setup of this paper.
In Sec.\ \ref{sec:simulation}, we present the simulation method to check the validity of the theoretical treatment.
Section \ref{sec:kinetic} is devoted to the explanation of the kinetic theoretical treatment.
Section \ref{sec:rheology} is the main part of this paper, where the rheology obtained from the simulation and the kinetic theory is shown.
In Secs.\ \ref{sec:discussion} and \ref{sec:conclusion}, we discuss and conclude our results.
In Appendix, we shortly summarize the explicit expression of the collision angle derived from the classical mechanics.


\section{Model and Setup}
Let us consider the three-dimensional system in which monodisperse particles (mass $m$ and diameter $\sigma$) are distributed randomly.
We assume that the position and velocity of $i$-th particle are given by $\bm{r}_i=(x_i, y_i, z_i)$ and $\bm{v}_i=(v_{i,x}, v_{i,y}, v_{i,z})$, respectively.
Here, the interaction between particles is assumed to be given by a harmonic potential
\begin{equation}
    U(r)=\frac{k}{2}(\sigma-r)^2\Theta(\sigma-r),
\end{equation}
where $k$ is the strength of the repulsion, $r$ is the interparticle distance, and $\Theta(x)$ is the step function.
We also consider the dissipative force proportional to the relative velocity between particles.
Therefore, the interparticle force between $i$-th and $j$-th particles is given by
\begin{equation}
    \bm{F}_{ij}=-\frac{\partial U(r_{ij})}{\partial \bm{r}_{ij}} -\zeta (\bm{v}_{ij}\cdot \hat{\bm{r}}_{ij})\hat{\bm{r}}_{ij}\Theta(\sigma-r_{ij}),
    \label{eq:force}
\end{equation}
where $\zeta$ is the dissipation ratio proportional to the relative velocity, $\bm{r}_{ij}\equiv \bm{r}_i-\bm{r}_j$, $r_{ij}\equiv |\bm{r}_{ij}|$, $\hat{\bm{r}}_{ij}\equiv \bm{r}_{ij}/r_{ij}$, and $\bm{v}_{ij}\equiv \bm{v}_i-\bm{v}_j$.
Here, the corresponding restitution coefficient is given by  $e=\exp[-\pi\zeta/(2mk-\zeta^2)^{1/2}]$.
In this paper, we mainly choose the dissipation rate as $\zeta=4.74\times10^{-2}\sqrt{mk}$, which corresponds to $e=0.9$.
We note that the choice of $e$ does not affect our results significantly.

\section{Simulation Method}\label{sec:simulation}
To check the validity of the theory explained in the next section, we perform the molecular dynamics simulation.
The detailed information is as follows:
We solve the equation of motion for each particle in terms of the Sllod equation \cite{Evans84, Evans}:
\begin{equation}
\begin{cases}
    \displaystyle \frac{d\bm{r}_i}{dt} = \frac{\bm{p}_i}{m}+\dot\gamma y_i \hat{\bm{e}}_x,\\
    \displaystyle \frac{d\bm{p}_i}{dt} = \sum_{j\neq i}\bm{F}_{ij}-\dot\gamma p_{i,y} \hat{\bm{e}}_x,\\
\end{cases}
\end{equation}
with the interparticle force \eqref{eq:force}, where $\dot\gamma$ is the shear rate, $\hat{\bm{e}}_x$ is the unit vector parallel to the $x$-direction, and $\bm{p}_i =(p_{i,x}, p_{i,y}, p_{i,z})=m(\bm{v}_i-\dot\gamma y_i \hat{\bm{e}}_x)$ is the peculiar momentum \cite{Evans84, Evans}.
We also adopt the periodic boundary condition in the $x$ and $z$-directions, and the Lees-Edwards boundary condition \cite{Lees72} in the $y$-direction.
In the following, we choose $m$, $\sigma$, and $k$ to nondimensionalize quantities to perform simulations.
The dimensionless time increment of the simulation is chosen as $\Delta t^*(\equiv \Delta t/(m/k)^{1/2})=0.01 \min(1, 1/\dot\gamma^*)$ with the dimensionless shear rate $\dot\gamma^*\equiv \dot\gamma/(k/m)^{1/2}$, which is sufficiently smaller than the collision duration and the characteristic time scale determined by the shear.
In this paper, we use $N=10^3$ particles and we fix the packing fraction as $\varphi=0.01$, which means that the linear length of the cubic system is chosen as $L=37.4\sigma$.

It is known that the rheology of dilute granular gases is characterized by the temperature $T\equiv (P_{xx}+P_{yy}+P_{zz})/(3n)$, the temperature difference $\Delta T\equiv (P_{xx}-P_{yy})/n$, and the shear stress $P_{xy}$ (or the viscosity $\eta \equiv -P_{xy}/\dot\gamma$) with the density $n$ \cite{Santos04, Hayakawa19, Takada18, Sugimoto20}.
For this purpose, we first measure the stress tensor as
\begin{equation}
    P_{\alpha\beta}=\frac{1}{L^3}\sum_i m v_{i,\alpha} v_{i,\beta},
\end{equation}
where $\alpha$ and $\beta$ indicate $x$, $y$, and $z$.
Using this, we can measure the flow quantities.
We note that another anisotropic temperature $\delta T \equiv (P_{xx}-P_{zz})/n$ is important in denser systems \cite{Hayakawa17, Saha17, Saha20, Takada20}.
However, this becomes equivalent to $\Delta T$ in dilute systems \cite{Hayakawa17, Takada20}.

\section{Kinetic Theory}\label{sec:kinetic}
In this section, we derive the flow curve of this system in terms of the kinetic theory.
In the kinetic theoretical treatment, the collision process is assumed to occur instantaneously.
Once we adopt the restitution coefficient $e$, the relationship between the pre- and post-collisional velocities ($\bm{v}_i^{\prime\prime}$ and $\bm{v}_i$, respectively) is given by
\begin{equation}
    \begin{cases}
    \displaystyle \bm{v}_1^{\prime\prime} = \bm{v}_1 - \frac{1+e}{2e}(\bm{v}_{12}\cdot \hat{\bm{k}})\hat{\bm{k}},\\
    \displaystyle \bm{v}_2^{\prime\prime} = \bm{v}_2 + \frac{1+e}{2e}(\bm{v}_{12}\cdot \hat{\bm{k}})\hat{\bm{k}},
    \end{cases}
\end{equation}
where $\hat{\bm{k}}$ is the unit vector from one particle to the other.

Now, let us consider the velocity distribution function $f(\bm{V},t)$, where $\bm{V}\equiv \bm{v}-\dot\gamma y \hat{\bm{e}}_x$ is the peculiar velocity.
In this case, the Boltzmann equation under shear is written as \cite{Santos04, Takada18, Chapman, Sugimoto20}
\begin{equation}
    \left(\frac{\partial}{\partial t} -\dot\gamma V_y \frac{\partial}{\partial V_x}\right)f(\bm{V},t)=J(\bm{V}|f),
    \label{eq:Boltzmann}
\end{equation}
with the collision operator $J(\bm{V}|f)$:
\begin{align}
    J(\bm{V}_1|f)
    &=\int d\bm{V}_2 \int d\hat{\bm{k}}
    \Theta(\sigma-b)\left|\bm{V}_{12}\cdot \hat{\bm{k}}\right|\nonumber\\
    &\hspace{1em}\times\left[e^{-2}\sigma_{\rm s}\left(\theta,V_{12}^{\prime\prime}\right) f\left(\bm{V}_1^{\prime\prime},t\right)
    f\left(\bm{V}_2^{\prime\prime},t\right)\right.\nonumber\\
    &\hspace{2.5em}\left.-\sigma_{\rm s}\left(\theta,V_{12}\right)
    f\left(\bm{V}_1,t\right)
    f\left(\bm{V}_2,t\right)\right],
\end{align}
where $\bm{V}_{12}\equiv \bm{V}_1 - \bm{V}_2$, $V_{12}\equiv |\bm{V}_{12}|$, and $\sigma_{\rm s}\equiv (1/2)(b/\sin(2\theta))|\partial b/\partial \theta|$ is the collision cross section determined from the relative speed and the collision angle $\theta$.
Here, the collision angle $\theta$ is given by \cite{Sugimoto20}
\begin{equation}
    \theta=\sin^{-1}\frac{b}{\sigma}+C_1F(\phi,\mathfrak{m})
    +C_2\Pi(a;\phi|\mathfrak{m})+C_3 \tan^{-1}\gamma + C_4,
    \label{eq:theta}
\end{equation}
where the quantities $\phi$, $\mathfrak{m}$, $a$, and $\gamma$ are functions of the impact parameter $b$ and the relative speed $v$ \cite{Chapman}.
We note that $F(\phi,\mathfrak{m})\equiv \int_0^\phi d\phi^\prime /(1-\mathfrak{m} \sin^2\phi^\prime)^{1/2}$ and $\Pi(a,\phi|\mathfrak{m})\equiv \int_0^\phi d\phi^\prime /\{(1-a\sin^2\phi^\prime)(1-\mathfrak{m}\sin^2\phi^\prime)^{1/2}\}$ are the elliptic integrals of the first and third kind, respectively \cite{Abramowitz}.
The detailed expressions of $C_1$, $C_2$, $C_3$, $C_4$, $\phi$, $\gamma$, $\mathfrak{m}$, and $a$ are listed in Appendix (see also Table I of Ref.\ \cite{Sugimoto20}).

\begin{figure}[htbp]
	\centering
	\includegraphics[width=\linewidth]{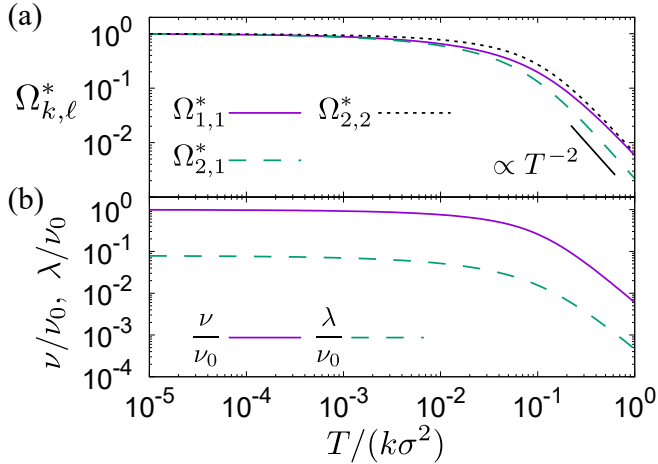}
	\caption{(Color online) (a) Temperature dependencies of the dimensionless Omega integrals $\Omega_{1,1}^*$ (solid line), $\Omega_{2,1}^*$ (dashed line), and $\Omega_{2,2}^*$ (dotted line).
	The guideline represents $T^{-2}$.
	(b) Temperature dependencies of the two frequencies $\nu$ (solid line) and $\lambda$ (dashed line) for $e=0.9$.}
	\label{fig:Omega_T}    
\end{figure}

By multiplying Eq.\ \eqref{eq:Boltzmann} by $mV_\alpha V_\beta$ and integrating over $\bm{V}$, we obtain the following evolution equation for the stress $P_{\alpha\beta}$ as
\begin{equation}
    \frac{\partial}{\partial t} P_{\alpha\beta}
    + \dot\gamma (\delta_{\alpha x}P_{y\beta}
    +\delta_{\beta x}P_{y\alpha})
    = -\Lambda_{\alpha\beta},
    \label{eq:stress_eq}
\end{equation}
where the right hand side of Eq.\ \eqref{eq:stress_eq} is defined as
\begin{equation}
    \Lambda_{\alpha\beta}
    \equiv -m \int d\bm{V}_1 V_{1,\alpha}V_{1,\beta}J(\bm{V}|f).
    \label{eq:Lambda}
\end{equation}
Unfortunately, the explicit form of this quantity is not known.
However, once we adopt Grad's approximation \cite{Hayakawa19, Hayakawa17, Takada18, Sugimoto20, Takada20}:
\begin{equation}
    f(\bm{V},t)\approx f_{\rm eq}(\bm{V},t)\left[1+\frac{m}{2T}\left(\frac{P_{\alpha\beta}}{nT}-\delta_{\alpha\beta}\right)V_\alpha V_\beta\right],
\end{equation}
with
\begin{equation}
    f_{\rm eq}(\bm{V},t)
    = n\left(\frac{m}{2\pi T}\right)^{3/2}\exp\left(-\frac{mV^2}{2T}\right),
\end{equation}
we can obtain closed equations.
It is noted that this approximation works well at least for hard-core systems even when the system is moderately dense \cite{Santos04, Hayakawa19, Hayakawa17, Takada18, Sugimoto20, Takada20}.
Then, under this approximation, we can rewrite Eq.\ \eqref{eq:Lambda} as \cite{Santos04, Hayakawa19, Hayakawa17, Takada18, Takada20}
\begin{equation}
    \Lambda_{\alpha\beta}
    =\nu (P_{\alpha\beta}-nT\delta_{\alpha\beta}) 
    + \lambda nT \delta_{\alpha\beta},
\end{equation}
where the diagonal and off-diagonal quantities $\nu$ and $\lambda$ are, respectively, given by
\begin{equation}
\begin{cases}
    \displaystyle \frac{\lambda}{\nu_0}
    = \frac{5}{12}(1-e^2)\Omega_{1,1}^*(T^*),\\
    \displaystyle \frac{\nu}{\nu_0} 
    = \frac{1+e}{4}\left[2(1-e)\Omega_{2,1}^*(T^*) + (1+e)\Omega_{2,2}^*(T^*)\right],\\
\end{cases}
\end{equation}
with the dimensionless temperature $T^*\equiv T/(k\sigma^2)$.
Here, we have introduced the frequency for the elastic hard-core system $\nu_0$ \cite{Santos04} as
\begin{equation}
    \nu_0 = \frac{16}{5}n\sigma^2 \sqrt{\frac{\pi T}{m}},
\end{equation} 
and the dimensionless Omega integral \cite{Chapman, Sugimoto20} 
\begin{align}
    \Omega_{k,\ell}^*
    &= C_{k,\ell}\int_0^\infty dy y^{2k+3}e^{-y^2}
    \int_0^1 db^* b^*\nonumber\\
    &\hspace{1em}\times
    \left\{1-(-1)^\ell \cos^\ell \left[2\theta\left(b^*, 2y\sqrt{T^*}\right)\right]\right\},
\end{align}
with $C_{k,\ell}=2$, $2/3$, and $1$ for $(k,\ell)=(1,1)$, $(2,1)$, and $(2,2)$, respectively.
Figure \ref{fig:Omega_T}(a) shows the temperature dependencies of the dimensionless Omega integrals.
In the low temperature limit, all quantities converge to unity, which means that the particles behave as hard-core gases.
In the high temperature limit, on the other hand, the quantities decrease as $\Omega_{k,\ell}^*\propto T^{*-2}$.
We also plot the temperature dependencies of $\nu$ and $\lambda$ for $e=0.9$ in Fig.\ \ref{fig:Omega_T}(b).
Because these quantities are written as the linear combination of $\Omega_{k,\ell}^*$, we can observe the same temperature dependencies.

Now, let us consider the rheology of this system.
From Eq.\ \eqref{eq:stress_eq}, we can obtain a set of dynamic equations:
\begin{equation}
\begin{cases}    
    \displaystyle \frac{dT}{dt} = -\frac{2\dot\gamma}{3n}P_{xy} - \lambda T,\\
    \displaystyle \frac{d\Delta T}{dt} = -\frac{2\dot\gamma}{n}P_{xy}-\nu\Delta T,\\
    \displaystyle \frac{dP_{xy}}{dt} =\dot\gamma n \left(\frac{1}{3}\Delta T-T\right) -\nu P_{xy}.
\end{cases}
    \label{eq:balance_eqs}
\end{equation}
In the steady state, we can obtain the temperature dependencies of these quantities as
\begin{equation}
\begin{cases}
    \displaystyle \dot\gamma=\nu\sqrt{\frac{3}{2}\frac{\lambda}{\nu-\lambda}},\quad
    \Delta T =\frac{3\lambda}{\nu}T,\\
    \displaystyle P_{xy}=-\frac{nT}{\nu}\sqrt{\frac{3}{2}\lambda(\nu-\lambda)},\quad
    \eta \equiv -\frac{P_{xy}}{\dot\gamma}= nT\frac{\nu-\lambda}{\nu^2},
\end{cases}
    \label{eq:flow_curve}
\end{equation}
respectively (see also Refs.\ \cite{Takada18, Hayakawa19}).
In the next section, we compare these results with those obtained from the simulations.

\section{Rheology}\label{sec:rheology}
\begin{figure}[htbp]
	\centering
	\includegraphics[width=\linewidth]{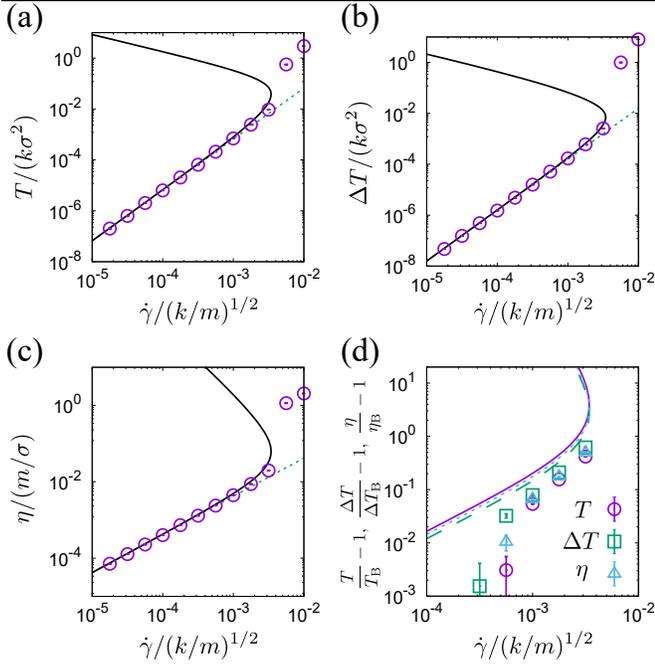}
	\caption{(Color online) Plots of (a) the temperature, (b) the temperature difference, and (c) the viscosity against the shear rate obtained from the theory (solid lines) and the simulations (open circles) for $e=0.9$.
	The dotted lines represent those for the hard-core limits \eqref{eq:Bagnold}.
	(d) The plots of the deviations of the quantities from the hard-core limits.
	The lines and marks represent the theoretical and numerical results, respectively.}
	\label{fig:flow_curve}    
\end{figure}

In this section, let us compare the results from the simulations and those from the kinetic theory.
Figure \ref{fig:flow_curve} shows the shear rate dependencies of (a) the temperature, (b) the temperature difference, and (c) the viscosity to those for hard-core gases, respectively.
Here, the low temperature limits are consistent with those for the hard-core limits (Bagnoldian) \cite{Santos04, Takada18}:
\begin{equation}
\begin{cases}
    \displaystyle T_{\rm B}
    = \frac{5\pi(2+e)}{432(1-e)(1+e)^2(3-e)^2}\frac{1}{\varphi}m\sigma^2\dot\gamma^2,\\
    \displaystyle \Delta T_{\rm B} 
    = \frac{25\pi(2+e)}{432(1+e)^2(3-e)^3}\frac{1}{\varphi^2}m\sigma^2 \dot\gamma^2,\\
    \displaystyle \eta_{\rm B}
    = \frac{5(2+e)}{72(1+e)^2(3-e)^3}\sqrt{\frac{5(2+e)}{3(1-e)}}\frac{1}{\varphi}\frac{m}{\sigma}\dot\gamma.
\end{cases}
    \label{eq:Bagnold}
\end{equation}
This is because the trajectories of the particles are coincide with those for hard-core particles.
As the shear rate increases, on the other hand, the deviations increase as shown in Fig.\ \ref{fig:flow_curve}(d), e.g., become approximately $10\%$ at $\dot\gamma \simeq 1\times 10^{-3} (k/m)^{1/2}$.
Interestingly, there is no theoretical solution for $\dot\gamma \gtrsim 3\times10^{-3} (k/m)^{1/2}$ while there exists another shear rate dependencies for $\dot\gamma \gtrsim 3\times10^{-3} (k/m)^{1/2}$ in the simulations (see panels (a)--(c) of Fig.\ \ref{fig:flow_curve}).
This occurs because our treatment of the inelasticity is not valid in this regime.
However, our theory works at least for $\dot\gamma\lesssim 2\times 10^{-2} (k/m)^{1/2}$, at which the temperature, the temperature difference, and the viscosity become approximately two times larger than those for the hard-core limits.

\begin{figure}[htbp]
	\centering
	\includegraphics[width=\linewidth]{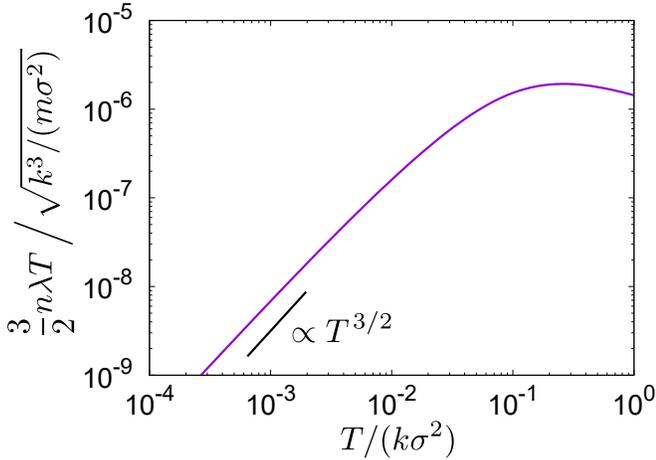}
	\caption{(Color online) Plot of the energy loss against the shear rate for $e=0.9$.
	The guideline represents $T^{3/2}$.}
	\label{fig:energy_loss}    
\end{figure}

Let us consider the reason why our kinetic theoretical treatment cannot predict steady states for $\dot\gamma \gtrsim 3\times10^{-3} (k/m)^{1/2}$.
Figure \ref{fig:energy_loss} shows the temperature dependence of the energy loss due to inelastic collisions ($3n\lambda T/2$) when we consider the energy balance equation (see also the first equation of Eqs.\ \eqref{eq:balance_eqs}).
In the low temperature regime, this energy loss coincides with that for the hard-core gases, where it is proportional to $T^{3/2}$.
As the shear rate increases, the increase of this loss decreases.
There exists a peak at $T\simeq 2.5\times10^{-1}k\sigma^2$ as shown in Fig.\ \ref{fig:energy_loss}, and then, the loss starts to decrease.
This means that the energy loss due to collisions cannot be balanced with the energy input by the shear, which is the origin of the absence of steady states.
It should be noted that the absence actually begins below this temperature, which is because the energy dissipation becomes smaller than that for hard-core gases as shown in Fig.\ \ref{fig:Omega_T}(b).


\section{Discussion}\label{sec:discussion}
\begin{figure}[htbp]
	\centering
	\includegraphics[width=\linewidth]{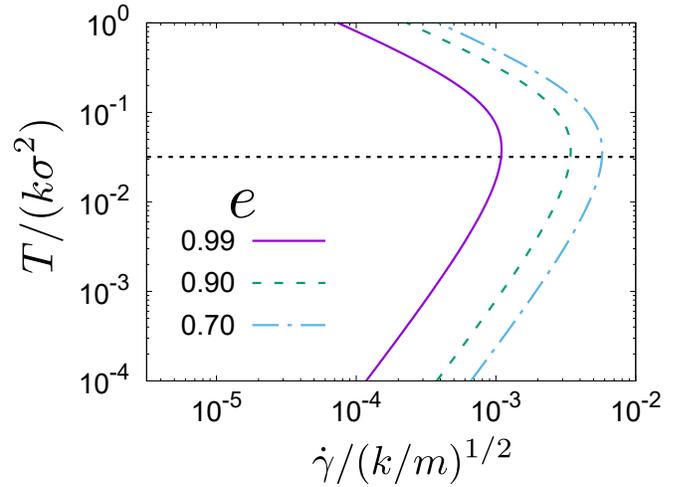}
	\caption{(Color online) Plots of the temperature against the shear rate for $e=0.99$ (solid line), $0.90$ (dashed line), and $0.70$ (dot--dashed line) obtained from the kinetic theory \eqref{eq:flow_curve}.
	The dotted line represents $T^*=0.1/\pi\simeq 3.2\times10^{-2}$.}
	\label{fig:T_shear_e}    
\end{figure}

As explained in the previous section, we can observe that the kinetic theory cannot reproduce the simulation results when the temperature is approximately larger than $T\simeq 4.0\times10^{-2}k\sigma^2$.
This can be understood as follows:
As explained in Ref.\ \cite{Sugimoto20}, the collision duration is approximately given by
\begin{equation}
    t_{\rm coll}\simeq \pi \sqrt{\frac{m}{2k}}.
\end{equation}
On the other hand, the characteristic time scale about the thermal velocity $v_{\rm T}=\sqrt{2T/m}$ becomes
\begin{equation}
    t_{\rm T} = \frac{\delta}{v_{\rm T}},\label{eq:t_T}
\end{equation}
where $\delta$ is the overlap between two colliding particles.
In the shear thinning regime ($T\gtrsim 4.0\times10^{-2}k\sigma^2$), $t_{\rm T}$ becomes smaller than $t_{\rm coll}$, i.e., $t_{\rm T}\lesssim t_{\rm coll}$.
Because the characteristic thermal time \eqref{eq:t_T} is also determined by the shear, this condition means that the particles are affected by the shear during the contact.
However, such an effect is not included in our kinetic theoretical treatment.
Therefore, this might be the origin of the discrepancy between the theory and the simulation for $T\gtrsim 4.0\times 10^{-2}k\sigma^2$.
Because the mean overlap is approximately given by $\delta\simeq 0.1\sigma$ for any $b$ (see Ref.\ \cite{Sugimoto20}), this condition becomes
\begin{equation}
    T\gtrsim \frac{0.1}{\pi}k\sigma^2\simeq 3.2\times10^{-2}k\sigma^2.
\end{equation}
This estimation works well as shown in Fig.\ \ref{fig:T_shear_e}, where the curve bends at around this temperature for any choice of $e$.
We note that the effect of shear is perturbatively included in the kinetic theory for hard-core systems \cite{Takada20}.
The extension of this method to our system is important, but this is our future work.

\section{Conclusion}\label{sec:conclusion}
In this paper, we have investigated the rheology of the inelastic soft-core granular gases under shear.
Once we adopt the combination of the collision angle of soft-core particles and the inelastic collision rule, the flow curve can be derived in terms of the kinetic theory.
We have found that the theory can reproduce the simulation results when the shear rate is not so large, while there is no steady state in the kinetic theory in the high shear regime.
This origin is also found that the energy dissipation due to inelastic collisions cannot be balanced with the energy input by the shear.

In this study, we focus on the rheology of dilute systems.
Of course, the analysis of denser systems is more important when we compare with experiments.
We believe that a similar procedure is available if we consider the Enskog equation, which considers the finite size effect of particles in the collision integral.
This extension is interesting and important, but this is our future work.

\begin{acknowledgment}
One of the authors (ST) thanks Hisao Hayakawa, Michio Otsuki, and Kuniyasu Saitoh for discussions.
This work is partially supported by the Grant-in-Aid of MEXT for Scientific Research (Grant No.\ JP20K14428).
\end{acknowledgment}

\appendix
\section{Detailed Expression of the Collision Angle $\theta$}
In this Appendix, we show the detailed expression of the collision angle $\theta$ in Eq.\ \eqref{eq:theta}.
Because the derivation is already explained in Ref.\ \cite{Sugimoto20}, we only summarize the result in Table \ref{fig:coeff_list}.
Here, we have introduced the complete elliptic integrals of the first and third kinds as $K(\mathfrak{m})\equiv F(\pi/2,\mathfrak{m})$ and $\Pi(a,\mathfrak{m})\equiv \Pi(a,\pi/2|\mathfrak{m})$, respectively \cite{Abramowitz}.
\begin{table}[htbp]
	\centering
	\caption{The detailed expressions of the quantities in Eq.\ \eqref{eq:theta}.}
 	\begin{tabular}{|l|}
 	\hline
 	$\displaystyle b^*= \frac{b}{\sigma}\quad \vline
    \quad v^*= \frac{v}{\sigma\sqrt{k/m}}$\\ \hline 
    $\displaystyle p = \frac{2-v^{*2}}{b^{*2}v^{*2}}\quad \vline
    \quad q = -\frac{4}{b^{*2}v^{*2}}\quad \vline
    \quad r = -\frac{q}{2}$\\ \hline
    $\displaystyle P = -\left(\frac{p^2}{3}+4r\right)\quad \vline
    \quad Q = -\frac{2}{27}p^3 -q^2 +\frac{8}{3}pr\quad \vline
    \quad \Delta = \left(\frac{Q}{2}\right)^2 + \left(\frac{P}{3}\right)^3$\\ \hline
    $\displaystyle \beta = 
    \begin{cases}
        \displaystyle -\frac{2p}{3}+\left(-\frac{Q}{2}+\sqrt{\Delta}\right)^{1/3}+\left(-\frac{Q}{2}-\sqrt{\Delta}\right)^{1/3} & (\Delta \ge 0)\\
        \displaystyle -\frac{2p}{3}+3\sqrt{-\frac{P}{3}}\cos\left\{\frac{1}{3}\cos^{-1}\left[-\frac{Q}{2}\left(-\frac{3}{P}\right)^{3/2}\right]\right\} & (\Delta < 0)
    \end{cases}$\\ \hline
    $\displaystyle D_1 = -\beta-2p+\frac{2q}{\sqrt{\beta}}\quad \vline
    \quad D_2 = -\beta-2p-\frac{2q}{\sqrt{\beta}}$\\ \hline
    $\displaystyle w_0 = -\frac{\sqrt{q^2+2\beta^2(p+\beta)}+q-2\beta}{\sqrt{q^2+2\beta^2(p+\beta)}-q+2\beta}$\\ \hline
    $\displaystyle \alpha_1 = \frac{q+\sqrt{q^2+2\beta^2(p+\beta)}}{2\beta}\quad \vline
    \quad \alpha_2 = \frac{q-\sqrt{q^2+2\beta^2(p+\beta)}}{2\beta}$\\ \hline
    $\displaystyle A_1 = \sqrt{-\frac{\beta^2 D_1}{\sqrt{q^2+2\beta^2(p+\beta)}-q-\beta^{3/2}}}$\\ \hline
    $\displaystyle A_2 = \sqrt{ \frac{\beta^2 D_2}{\sqrt{q^2+2\beta^2(p+\beta)}-q+\beta^{3/2}}}$\\ \hline
    $\displaystyle \mathfrak{m} =
    \begin{cases}
        \displaystyle \frac{A_2^2}{A_1^2} & (D_1\ge 0) \\ 
        \displaystyle \frac{A_2^2}{A_1^2+A_2^2} & (D_1<0)
    \end{cases}\quad \vline
    \quad 
    \phi = 
    \begin{cases}
        \displaystyle \sin^{-1}\frac{w_0}{A_2} & (D_1\ge 0)\\
        \displaystyle \cos^{-1}\frac{w_0}{A_2} & (D_1<0)
    \end{cases}$\\ \hline
    $\displaystyle 
    \gamma =
    \begin{cases}
        \displaystyle \sqrt{\frac{\left(A_1^2-1\right)\left(A_2^2-w_0^2\right)}{\left(1-A_2^2\right)\left(A_1^2-w_0^2\right)}} & (D_1\ge 0)\\
        \displaystyle \sqrt{\frac{\left(A_1^2+1\right)\left(A_2^2-w_0^2\right)}{\left(1-A_2^2\right)\left(A_1^2+w_0^2\right)}} & (D_1<0)
    \end{cases}\quad \vline 
    \quad 
    a =
    \begin{cases}
        \displaystyle A_2^2 & (D_1\ge 0)\\
        \displaystyle -\frac{A_2^2}{1-A_2^2} & (D_1<0)
    \end{cases}$\\ \hline
    $\displaystyle 
    C = 
    \begin{cases}
        \displaystyle \frac{2\sqrt{A_1A_2}}{\left(D_1D_2\right)^{1/4}} & (D_1\ge 0)\\
        \displaystyle \frac{2\sqrt{A_1A_2}}{\left(-D_1D_2\right)^{1/4}} & (D_1<0)
    \end{cases}\quad \vline 
    \quad
    C_1 =
    \begin{cases}
        \displaystyle -\frac{\alpha_2}{A_1} & (D_1\ge 0)\\
        \displaystyle \frac{\alpha_2}{\sqrt{A_1^2+A_2^2}} & (D_1<0)
    \end{cases}$\\ \hline
    $\displaystyle 
    C_2 =
    \begin{cases}
        \displaystyle -\frac{\alpha_1-\alpha_2}{A_1} & (D_1\ge 0)\\
        \displaystyle \frac{\alpha_1-\alpha_2}{\left(1-A_2^2\right)\sqrt{A_1^2+A_2^2}} & (D_1<0)
    \end{cases}$\\ \hline
    $\displaystyle C_3 =
    \begin{cases}
        \displaystyle \frac{\alpha_1-\alpha_2}{\sqrt{\left(A_1^2-1\right)\left(1-A_2^2\right)}} & (D_1\ge 0)\\
        \displaystyle \frac{\alpha_1-\alpha_2}{\sqrt{\left(A_1^2+1\right)\left(1-A_2^2\right)}} & (D_1<0)
    \end{cases}$\\ \hline
    $\displaystyle 
    C_4 =
    \begin{cases}
        \displaystyle -C_1K(\nu) - C_2 \Pi(a,\nu) & (D_1\ge 0)\\
        0 & (D_1<0)
    \end{cases}$\\ \hline
    $\displaystyle C_1 = C C_1^\prime \quad \vline
    \quad C_2 = C C_2^\prime \quad \vline
    \quad C_3 = C C_3^\prime \quad \vline
    \quad C_4 = C C_4^\prime$\\
 	\hline
	\end{tabular}
	\label{fig:coeff_list}
\end{table}


\end{document}